\documentclass[runningheads]{llncs}
\usepackage{geometry}
\usepackage{cite}
\usepackage{censor}
\usepackage{textcomp}
\usepackage[utf8]{inputenc} % allow utf-8 input
\usepackage[T1]{fontenc}    % use 8-bit T1 fonts
\usepackage{hyperref}       % hyperlinks
\usepackage{url}            % simple URL typesetting
\usepackage{booktabs}       % professional-quality tables
\usepackage{amsfonts}       % blackboard math symbols
\usepackage{nicefrac}       % compact symbols for 1/2, etc.
\usepackage{microtype}      % microtypography
\usepackage{xcolor}         % colors
\usepackage{amsmath,amssymb,amsfonts}
\usepackage{algorithmic}
\usepackage{graphicx}
\usepackage{blindtext}
\usepackage{textcomp}
\usepackage[notransparent]{svg}
\usepackage{verbatim}
\usepackage{color,soul}
\usepackage{graphics}
\usepackage{listings}
\usepackage{multirow}
\usepackage{tabularx}
\usepackage{amssymb}
\usepackage[toc,page]{appendix}
\usepackage{mwe}%
\usepackage{todonotes}

\def\BibTeX{{\rm B\kern-.05em{\sc i\kern-.025em b}\kern-.08em
    T\kern-.1667em\lower.7ex\hbox{E}\kern-.125emX}}

\usepackage{xcolor}
\usepackage{xspace}
\newif\ifdraft
\drafttrue

\ifdraft
  \newcommand{\jhanote}[1]{{\textcolor{red}{ ***Shantenu: #1 }}\xspace}
\else
 \newcommand{\jhanote}[1]{}
 \fi
    
\begin{document}

\title{Deep RC: A Scalable Data Engineering and Deep Learning Pipeline}

\author{Arup Kumar Sarker\inst{1, 3} \and
Aymen Alsaadi\inst{2}\and
Alexander James Halpern \inst{1}\and
Prabhath Tangella\inst{1}\and
Mikhail Titov\inst{4} \and
Niranda	Perera\inst{7} \and
Mills Staylor\inst{1} \and
Gregor von Laszewski\inst{3} \and
%Andre Merzky\inst{2} \and
Shantenu Jha\inst{2,5,6} \and
Geoffrey Fox\inst{1,3} 
}
\authorrunning{Sarker et al.}
\institute{Department of Computer Science, University of Virginia, Charlottesville, VA 22904 \email{\{djy8hg, upy9gr, nww7sm, vxj6mb\}@virginia.edu} 
\and
Rutgers-New Brunswick, NJ, 08901-8554
\email{ \{aymen.alsaadi, shantenu.jha\}@rutgers.edu} \and
Biocomplexity Institute and Initiative, Town Center Four, 994 Research Park Boulevard Charlottesville, VA 22911
\email{laszewski@gmail.com}
\and
Brookhaven National Laboratory, Upton, NY
\email{mtitov@bnl.gov} 
\and
Princeton Plasma Physics Laboratory, Princeton, NJ
\and
Princeton University, Princeton, NJ
\and
Nvidia Corporation, Santa Clara, CA
\email{niranda@niranda.dev} }

%\and
%Brookhaven National Laboratory, 98 Rochester St, Upton, NY 11973
%\email{ \{okilic, mtitov\}@bnl.gov} 
%\and 
%Voltron Data, Mountain View, 650 Castro St, CA
%\email{niranda.perera@gmail.com}
%}
%

\begin{comment}
\author{\IEEEauthorblockN{\censor{Given Name Surname}}
\IEEEauthorblockA{\censor{\textit{dept. name of organization (of Aff.)}} \\
\textit{\censor{name of organization (of Aff.)}}\\
\censor{City, Country} \\
\censor{email address or ORCID}}
\and
\IEEEauthorblockN{Given Name Surname}
\IEEEauthorblockA{\textit{dept. name of organization (of Aff.)} \\
\textit{name of organization (of Aff.)}\\
City, Country \\
email address or ORCID}
\and
\IEEEauthorblockN{Given Name Surname}
\IEEEauthorblockA{\textit{dept. name of organization (of Aff.)} \\
\textit{name of organization (of Aff.)}\\
City, Country \\
email address or ORCID}
\and
\IEEEauthorblockN{Given Name Surname}
\IEEEauthorblockA{\textit{dept. name of organization (of Aff.)} \\
\textit{name of organization (of Aff.)}\\
City, Country \\
email address or ORCID}
\and
\IEEEauthorblockN{\censor{Gregor von Laszewski}}
\IEEEauthorblockA{\textit{\censor{Biocomplexity Institute and Initiative}} \\
\textit{\censor{University of Virginia}}\\
\censor{Charlottesville, VA, 22911, USA} \\
\censor{laszewski@gmail.com} \\
\censor{https://orcid.org/0000-0001-9558-179X}}
\and
\IEEEauthorblockN{Given Name Surname}
\IEEEauthorblockA{\textit{dept. name of organization (of Aff.)} \\
\textit{name of organization (of Aff.)}\\
City, Country \\
email address or ORCID}
}
\end{comment}

\maketitle
\begin{abstract}
Significant obstacles exist in scientific domains including genetics, climate modeling, and astronomy due to the management, preprocess, and training on complicated data for deep learning. Even while several large-scale solutions offer distributed execution environments, open-source alternatives that integrate scalable runtime tools, deep learning and data frameworks on high-performance computing platforms remain crucial for accessibility and flexibility. In this paper, we introduce Deep Radical-Cylon(RC), a heterogeneous runtime system that combines data engineering, deep learning frameworks, and workflow engines across several HPC environments, including cloud and supercomputing infrastructures. Deep RC supports heterogeneous systems with accelerators, allows the usage of communication libraries like \texttt{MPI}, \texttt{GLOO} and \texttt{NCCL} across multi-node setups, and facilitates parallel and distributed deep learning pipelines by utilizing Radical Pilot as a task execution framework. By attaining an end-to-end pipeline including preprocessing, model training, and postprocessing with 11 neural forecasting models (PyTorch) and hydrology models (TensorFlow) under identical resource conditions, the system reduces 3.28 and 75.9 seconds, respectively. The design of Deep RC guarantees the smooth integration of scalable data frameworks, such as Cylon, with deep learning processes, exhibiting strong performance on cloud platforms and scientific HPC systems. By offering a flexible, high-performance solution for resource-intensive applications, this method closes the gap between data preprocessing, model training, and postprocessing.
\end{abstract}

%\pagestyle{plain}
%\keywords{HPC \and BSP \and Cylon \and ETL \and MPI \and UCX \and RP \and BM \and SPMD \and MPMD}

%
%\input{tex/intro.tex}

\section{Introduction}\label{sec:introduction}
It is a significantly complex and challenging task to manage and prepare massive data for deep learning in big data science which introduces additional difficulty in model training causing an impact on fields such as hydrology, genomics, climate modeling and astronomy ~\cite{McKenna2010The}.  Furthermore, the integration of data from diverse sources not only introduces significant heterogeneity and multifaceted structure but also manifests complex interdependencies among parameters, thereby exacerbating the challenges associated with the bare metal process. The sheer magnitude and complexity of these enormous datasets place significant processing limitations on even the most advanced computing infrastructures, frequently resulting in computational bottlenecks that impede effective data analysis and interpretation. \cite{sarker2024radical}. %Moreover, the sheer volume of data can make storing and transferring data across systems challenging. 
Google Pathways\cite{barham2022pathways, introduc61:online} and OneFlow\cite{yuan2021oneflow} address some aspects of these challenges. But the design and distributed runtime of those systems are black-box and there is no way to measure the comparative results of a heterogeneous data pipeline. We aim to develop a unified approach combining data engineering and deep learning frameworks with diverse execution capabilities, which can be deployed on various HPC platforms, including cloud systems and HPCs.
%Integrating scalable HPC runtime tools with data frameworks is vital for creating scalable solutions for processing large-scale simulations, modeling, and machine learning. 

To address this, Cylon\cite{widanage2020high} offers foundational frameworks for data engineering on scalable HPC machines. DASK\cite{rocklin2015dask} and SPARK\cite{zaharia2016apache} Dataframe can be alternative to Cylon.  However, Cylon outperforms both in multiple scaling operations\cite{perera2023depth, widanage2020high, abeykoon2020data} although Cylon lacks optimization for efficient resource use and does not support a heterogeneous data pipeline. On the other hand, our previous work Radical-Cylon \cite{sarker2024radical}, a task-based architecture, enables Cylon to interact and operate with different HPC platforms seamlessly, shielding Cylon from heterogeneous configurations of different HPC platforms that support data engineering pipelines. However, it lacked support for deep learning execution and the use of heterogeneous resources. To execute the deep learning job by using the cylon dataframe, we had to preprocess the dataframe separately, and then use it as a dataset for the deep learning model. Moreover, It has limitations supporting GPUs and a combination of CPU-GPU execution based on the platform. On the other hand, RAY\cite{moritz2018ray} provides a distributed runtime that can be an alternative to RADICAL-Pilot \cite{merzky2022radical} as a workflow engine. However, Ray's GPU support is restricted to scheduling and reservations, whereas Radical pilot has the flexibility to control underlying hardware resources for scheduled tasks. We aim to enhance support by introducing Deep Radical-Cylon(RC), a heterogeneous runtime system with accelerators, enabling frameworks like Cylon with Pytorch and Tensorflow to leverage heterogeneous execution through RADICAL-Pilot and handle data pipelines from data engineering tasks with cylon-distributed data frame to downstream deep learning training and inferencing jobs. 
%Rapids-cuDF is highly optimized for NVidia GPUs. GPU Cylon is in the pipeline for scaling operations and can be a part of Cylon. 
%\TODO{at this time I am lost what is done and what is done} We introduce Deep Radical-Cylon (RC), a heterogeneous runtime system with a distributed and parallel data structure that uses RADICAL-Pilot to perform deep learning and Cylon tasks in a single pipeline.
%\TODO{this sunds now repeated} By constructing a data loader with a cylon-distributed data frame that feeds downstream deep learning training and inferencing tasks, it guarantees a full bridge \TODO{what is a bridge}. 
By separating resource management from the application layer, deep RC makes it possible for a job to execute on any HPC platform without the need for code rewrites or changes. This makes development easier and produces a system that is more adaptable and loosely integrated frameworks. In comparison to bare-metal deep Learning execution, this method achieves competitive performance with little overhead by enabling the usage of heterogeneous communicators across numerous nodes.
%\TODO{contradicts eralier statement wher you use sane. Arup: Fixed} 
%This combined functionality fosters collaboration and drives innovation within the open-source scientific research community. 
It offers enhanced functionalities such as distributed data pretreatment and post-processing inside a unified pipeline, accommodating heterogeneous data and model executions. Deep RC has been evaluated with 11 NeuralForecast \cite{olivares2022library_neuralforecast} models for PyTorch and hydrology models \cite{he2024science} for TensorFlow with inferencing jobs and reduces 3.28 and 75.9 seconds respectively.

\begin{figure}[htpb]
    \begin{center}
    \includesvg[width=0.55\linewidth]{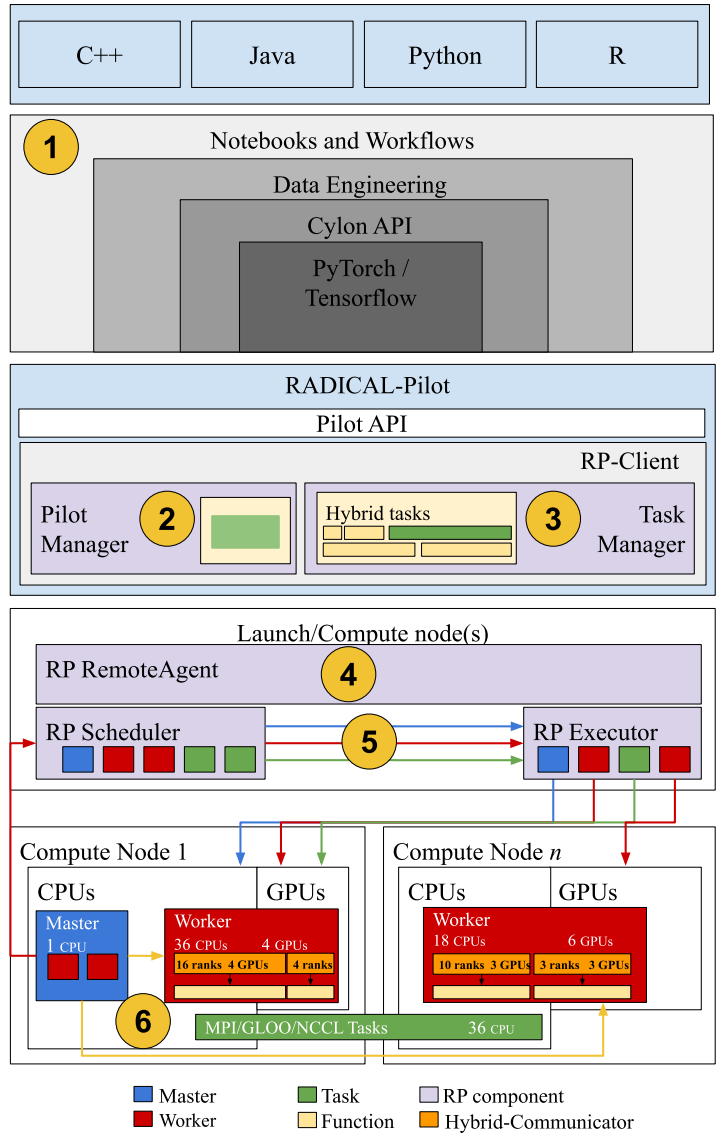}

    \caption{Deep RC design incorporating Radical-Cylon \cite{sarker2024radical} Architecture with GLOO and NCCL as communication framework. A modular design with dependent components. Segregated independent module with top-down flow from cross-platform to hardware resources. Application pipeline is incorporated with Cylon APIs and PyTorch and TensorFlow with proposed Deep RC Bridge in Fig.-\ref{fig:deep_cylon_design}  }
    \label{fig:cylon_rp_design}
    \end{center}
\end{figure}

\section{Design and Implementation}
\label{sec:design}

\subsection{Cylon}

%By establishing a link between reliable data processing pipelines and AI/ML frameworks like PyTorch ~\cite{pytorch2019} and TensorFlow~\cite{tensorflow2015}, Cylon's structure marks a substantial advancement in the field of data engineering. \TODO{unsubstantiated claim and only belongs in conclusion, this is now advertisement} Through seamless interaction with various data formats and systems, this integration exemplifies a clear vision of integrating data engineering and AI/ML workflows. \TODO{and what is the vision} 
To improve performance and lower the complexity of distributed data operations, Cylon is designed to optimize Extract, Transform, and Load (ETL) procedures \cite{widanage2020high}. At the heart of Cylon’s design lies a sophisticated core framework centered on a table abstraction that represents structured data. 
%\TODO{so ther eis a limit that it can not do unstructured data, is that needed for AI? maybe instead focus on which type of experiments can be run with structured data. Arup: We focused on structured data.}
This abstraction ensures a cooperative solution to distributed computing difficulties 
%\TODO{difficulties arising from using distributed compute resources, by the way why one needs distributed is not introduced} 
by allowing individual ranks or processes to handle data partitions jointly. Local and distributed operations are intended to be balanced by the Python and C++-based interface. Tasks involving only locally accessible data are handled by local operators, guaranteeing prompt and effective completion. Distributed operators(e.g. shuffle, gather, reduce, etc.), on the other hand, use sophisticated network capabilities to carry out intricate operations that call for inter-process communication, like data collecting, reduction, and shuffling.
%\TODO{Cylon provides a wide variety of data manipulation operatiors. THis includes operators working on local resources but also operators that work on distributed resources. Using these operators in a program simplifies the data processing as they provide high level abstractions that are readily reusable. Examples of such operators are ... and are discussed in xyz in more detail}

Cylon incorporates network-level 
%\TODO{you need to describe the difference between functional and network level operators in more detail. Are network related things actually operators or should they be somethings else.  Maybe explain what an operator actually is.Arup: It would be a good addition, but those does not have much more impact down those paper}
operations based on communication protocols such as TCP and Infiniband to handle the inherent challenges of distributed programming. Heterogeneous data transmission is made possible by these protocols' support for several communication abstraction frameworks, including MPI, UCX\cite{shamis2015ucx}, and GLOO\cite{gloo:online}. This channel abstraction 
%\TODO{now you switch to channel abstriction instead of using network operator and its not sufficiently explained what the difference between the two are. Arup: That is not part of this paper. We cited those paper} 
technique is essential for improving performance in a variety of hardware settings and simplifying communication across dispersed processes. High-efficiency operations like shuffle and reduce that depend on these abstractions allow for scalable and smooth distributed workflows\cite{shan2022hybrid, perera2023supercharging}.

%\TODO{I think you need a better description about ooperators and go from high level to low level. Arup: I have made some changes to create all operator descriptions. But Cylon in depth discussion is too broad.}

Cylon's data model is made compatible and interoperable by using Apache Arrow's Columnar Format as its foundation. Because of this, Cylon may easily interact with a wide range of open-source frameworks and libraries, guaranteeing compatibility and seamless data flow across the broader data ecosystem. This foundation encourages innovation in the creation of high-performance applications while supporting a consistent and effective approach to data processing. Cylon's architecture combines sophisticated communication techniques, local and distributed operators, and strong data models to create a high-performance distributed framework \cite{perera2023depth}. Additionally, it is versatility and useful in coordinating intricate workflows across disparate computer resources are highlighted by its abstraction under RADICAL-Pilot.

\subsection{Radical-Cylon}

%We published Radical-Cylon \cite{sarker2024radical} for heterogeneous data engineering jobs that seamlessly integrates Cylon and RADICAL-Pilot. It's a flexible runtime system to effectively manage concurrent and diverse workloads while meeting a variety of processing demands is its main strength. RADICAL-Pilot makes it easier to deploy complicated workflows across a variety of HPC resources by abstracting resource management through a flexible pilot-based approach.  The PilotManager, TaskManager, and RemoteAgent are the three primary parts of the system. Together, these elements provide a simplified execution environment. The PilotManager is responsible for managing the lifecycle of the pilot, a placeholder that secures and organizes resources on an HPC system. It ensures that resource requests and allocations are managed effectively by operating on user-accessible resources, such as a local computer or a cluster's login node. Tasks are the applications or functions that are run on the pilot resources, and the TaskManager manages their lifetime ~\cite{merzky2022radical}. It makes sure that tasks are scheduled and executed optimally, coordinating closely with the PilotManager to maximize resource utilization.
We published Radical-Cylon \cite{sarker2024radical} for heterogeneous data engineering jobs seamlessly integrating Cylon and RADICAL-Pilot (RP). RP is a flexible runtime system that effectively manages the concurrent execution of highly heterogeneous workloads on diverse heterogeneous resources. RP makes deploying complicated workloads across various HPC resources easier by abstracting resource management through a flexible pilot-based approach. The PilotManager, TaskManager, and RemoteAgent are the three primary parts of the system that provide a simplified execution environment on HPC resources. The PilotManager is responsible for managing the lifecycle of the pilot, a placeholder that acquires and manages the computes resources. It ensures that resource requests and allocations are managed effectively by operating on user-accessible resources, such as a local computer or a cluster's compute nodes. Tasks represent the user applications, which can be executables or functions managed by TaskManager~\cite{merzky2022radical}. The TaskManager works closely with the PilotManager to ensure efficient and effective task management and scheduling to maximize resource utilization.

When installed on an HPC system's computing nodes, the RemoteAgent sets up the execution environment and controls task execution. It guarantees smooth communication between the workload and the resources allotted, enabling RADICAL-Pilot to effectively carry out activities across dispersed nodes. Even in the most taxing computational situations, RADICAL-Pilot can provide high throughput and scalability thanks to its decentralized yet integrated design. 
%RADICAL-Pilot has proven to be able to run up to one million independent tasks simultaneously over one thousand nodes with low overhead using its pilot abstraction architecture. 
It supports a diverse range of workloads, including  MPI/GLOO/NCCL-based tasks, as well as single-threaded, multi-threaded, and multi-core applications. By offering such flexibility, RADICAL-Pilot caters to a variety of use cases, from large-scale simulations to data-intensive workflows, making it a vital tool for researchers and engineers working on HPC platforms~\cite{merzky2022radical}.%Utilizing their respective native Application Programming Interfaces (APIs), Radical-Cylon is an integrated framework that seamlessly integrates Cylon and RADICAL-Pilot. The distributed runtime in this system, RADICAL-Pilot, controls how Cylon-generated tasks are carried out.
By using a loosely coupled approach, the integration makes full use of both systems' capabilities via the RADICAL-Pilot API and eliminates the need for further integration plugins. Without needing any changes to Cylon activities, this design enables Cylon to leverage the heterogeneous runtime characteristics of RADICAL-Pilot, specifically, its capacity to generate and control MPI communicators ~\cite{sarker2024radical}.

\subsection{Design}
\begin{comment}
Cylon and RADICAL-Pilot are two isolated systems offering different functionalities and capabilities. The integration design advocates the loosely coupled approach where both systems work independently of each other, with minimal dependencies and interactions, while benefiting from each other's capabilities.

The integrated design of RADICAL-Pilot and Cylon is shown in Fig.~\ref{fig:cylon_rp_design} where Cylon is plugged as a top-level component to send different types of Cylon tasks (functions or executables) to RADICAL-Pilot to execute on HPC resources. The main communication point between both systems is their native APIs, as both systems offer flexible and simple Python-based interfaces. 

Cylon and RADICAL-Pilot loosely coupled integration can be easily scaled out, expanded, or contracted to meet changing demands~\cite{de_dynamio_2007}. Flexibility-wise, both systems are developing rapidly and might introduce new fundamental changes in the design or implementation, such as adding or removing new system components. Further, any changes in both systems do not necessarily require changes to the other system's components and would not affect the existing integration as there are no direct dependencies between the integrated systems. From a fault tolerance perspective, the integration approach of Cylon and RADICAL-Pilot is more resilient, as failures in one system or component do not affect the entire system. Failure of any component can be isolated and contained, allowing the rest of the system to continue receiving and executing tasks.    
\end{comment}

Deep learning model execution, Cylon and RADICAL-Pilot are three distinct systems that serve complementary purposes, each offering specialized functionalities and capabilities. The loosely linked architecture that underpins the end-to-end pipeline enables them to function independently while utilizing one another's advantages. This strategy minimizes dependencies and maximizes modularity by ensuring that neither system is firmly tied to the other. When workloads or computing demands change, the system can be scaled out, enlarged, or reduced. Furthermore, the absence of direct dependencies guarantees that any changes made to either system's existing components or the addition of new ones won't interfere with the integrated framework. Without requiring changes to the other system, each system can implement updates, add or remove components, or rethink essential features.

Cylon is a high-level component in this pipeline that is in charge of developing and overseeing tasks like executables or data engineering functions. RADICAL-Pilot is utilized as a workflow engine for execution on HPC resources after these tasks are sent to a deep learning framework as an input dataframe for training and inferencing jobs ~\cite{de_dynamio_2007}. The native APIs of the three systems, which both offer straightforward and adaptable Python-based interfaces, enable communication between them. Task handoff and execution are made possible by this architecture, which guarantees a smooth connection between the two without the need for intricate intermediary layers. For instance, RADICAL-Pilot schedules these operations to optimize the flow of data to CPUs or GPUs, while Cylon can preprocess big datasets into forms compatible with PyTorch's ~\texttt{DataLoader} ~\cite{pytorch2019}  or TensorFlow's ~\texttt{tf.data.Dataset} ~\cite{tensorflow2015}.  %By using native APIs, developers may provide scalability and extension while preserving integration clarity and simplicity. 
%The Deep RC's loosely linked architecture provides notable benefits in terms of adaptability and scalability. When workloads or computing demands change, the system can be scaled out, enlarged, or reduced. Furthermore, the absence of direct dependencies guarantees that any changes made to either system's existing components or the addition of new ones won't interfere with the integrated framework. Without requiring changes to the other system, each system can implement updates, add or remove components, or rethink essential features.

In terms of fault-tolerance, this integration is very robust. The overall stability of the framework is maintained since failures in one system or its components do not ripple into the other system. For example, RADICAL-Pilot is unaffected and can carry on with other tasks even if a task fails during the model training phase. Similarly, the deep learning framework's task management capabilities are unaffected by any errors or resource limitations in RADICAL-Pilot. The system can recover gracefully and carry on even in the face of adversity thanks to this isolation of failures. The design philosophy behind the pipeline exemplifies a robust and forward-looking approach to creating modular, scalable, and fault-tolerant frameworks. Through this interface, data engineering and deep learning workflows on HPC platforms are executed with flexibility and resilience, utilizing the benefits of both systems while preserving their independence. This architecture guarantees flexibility for upcoming advancements in both systems in addition to meeting present computing demands.

\subsection{Implementation}

Deep Radical-Cylon(RC) is a unified system that facilitates seamless communication between Cylon, RADICAL-Pilot, and deep learning frameworks through their Python APIs. To define, organize, and carry out deep learning jobs across many HPC systems, the architecture makes use of RADICAL-Pilot as the main interface. A \texttt{RadicalPilot.TaskDescription} object, which outlines resource requirements such CPUs, GPUs, and memory allocations, is used to represent each deep learning task. When specifying the computational requirements of each activity, this structured representation guarantees accuracy and clarity. 

In order to instruct the PilotManager to generate a Pilot object with the necessary resources at initialization, Deep RC uses RADICAL-Pilot. On HPC systems, this Pilot acts as a stand-in for resource management. At the same time, RADICAL-Pilot creates a TaskManager that is in charge of sending deep learning tasks, like model training or inference, as well as Cylon-specific tasks to be carried out on distant computing resources. The system works in tandem with the HPC resource manager, guaranteeing efficient coordination between task execution and resource acquisition. The RemoteAgent is deployed on the compute nodes by RADICAL-Pilot after the necessary resources have been assigned. RADICAL-Pilot deploys a multi-node master-worker execution environment, which offers the capabilities to dynamically construct MPI-Communicators to run heterogeneous NCCL/GLOO/MPI operations on several nodes at once (Fig.-\ref{fig:cylon_rp_design}). Such an approach allows the efficient execution of concurrent AI and HPC tasks, which frequently call for unique communication patterns among a subset of resources, to benefit greatly from this characteristic.

%The RAPTOR subsystem, an abstraction of the master-worker paradigm, is started by the RemoteAgent, which also bootstraps the execution environment. RAPTOR is made to run heterogeneous NCCL/GLOO/MPI operations on several nodes at once (Fig.-\ref{fig:cylon_rp_design}). This design offers the special ability to dynamically build communicators during runtime, in contrast to traditional pilot systems. Deep learning tasks, which frequently call for unique communication patterns among a subset of resources, benefit greatly from this characteristic.

%The RemoteAgent's scheduler sends a queue of Cylon and deep learning tasks to the master while both the worker and master processes of RAPTOR are running.
The RemoteAgent's scheduler sends a queue of Cylon and deep learning tasks to the master(s), and ultimately, each master distributes the tasks to the worker(s).
%The employees are then given certain tasks to complete.
Based on the particular resource requirements of the task, a worker isolates a set of parallelism upon receiving it. The task's communicator is then built and delivered at runtime, allowing for effective execution and task-specific communication. The master process gathers the outcomes of tasks as they are finished and sends them back to the TaskManager. Thanks to this organized feedback loop, all job results are centralized and made accessible for additional processing or analysis. RP's dynamic and adaptable features enable the integration of Cylon and deep learning frameworks within RADICAL-Pilot, resulting in a highly scalable and effective solution for carrying out intricate processes on HPC platforms. This architecture is an effective tool for large-scale data engineering and deep learning execution since it not only improves resource usage but also simplifies the execution of heterogeneous activities.

%CYLON and RADICAL-Pilot are two isolated systems offering different functionalities and capabilities. 
The ability to use GLOO~\cite{gloo:online} and NCCL~\cite{nccl:online} as the communication backend for distributed operations is one of the enhanced advantages of the suggested architecture on Deep RC. The design encourages a loosely connected approach where both systems benefit from one another's advantages while functioning separately with few interactions and dependencies. We provide a distributed data loader for the Cylon data frame so that the preprocessed Cylon dataframe can be used as input for the deep learning system. We introduce two kinds of bridges: Data Bridge, which connects Cylon to the deep learning framework, and System Bridge, which manages the resources and flow from Cylon to RADICAL-Pilot(RP). Deep RC is the name given to the complete system.

\begin{figure}[htbp]
\begin{center}
    \includesvg[width=0.95\linewidth]{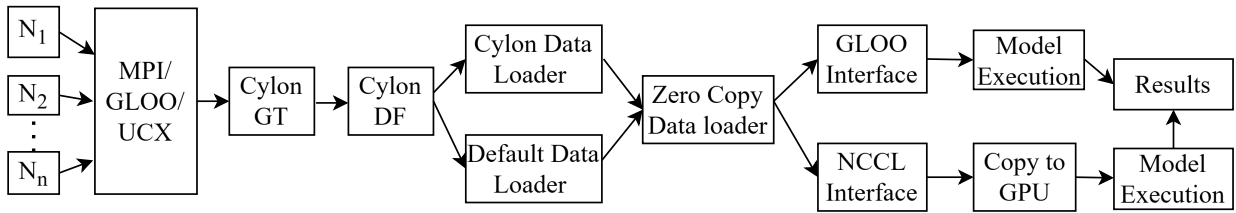}
\end{center}
     \caption{Deep RC Bridge. From distributed data pre-processing to deep learning model execution. Cylon distributed Tasks are scheduled and executed on CPUs with MPI/GLOO/UCX communication frameworks. }
    \label{fig:deep_cylon_design}
\end{figure}

In Fig.-~\ref{fig:deep_cylon_design}, we demonstrated the Deep RC bridge, which allows data to be preprocessed using Cylon distributed dataframes that operate on top of MPI/UCX/GLOO and produce a Cylon Global Table (GT). The global table, which may be zero-copied and translated to pandas and other data frame formats, is used to create the distributed Cylon dataframe. Before batches are loaded into memory, data transformations and augmentation are frequently implemented. This guarantees that the training process is not slowed down by preprocessing operations like shrinking photos, standardizing values, or applying random augmentations. %The transformations are typically performed on the CPU before sending the processed data to the GPU. Additionally, data collation ensures that different samples within a batch are correctly formatted, especially when working with variable-length sequences or multi-modal inputs.
Our custom data loader was designed to inherit the functionality of the default PyTorch/TensorFlow data loader, enabling it to efficiently traverse the dataset as required. A batch of \texttt{train\_features} and \texttt{train\_labels} are returned at the end of each iteration. Since it requires no extra memory to build training and verification datasets, we dubbed it a zero-copy data loader. The zero-copy data loader uses several workers to retrieve and preprocess data in simultaneously, effectively managing data loading in the Deep RC pipeline. It divides the workload among several subprocesses, each of which is in charge of loading a section of the dataset, rather than depending on a single process to load data sequentially. By preventing training bottlenecks, this parallelism makes sure the model gets data as soon as possible. The main objective is to minimize waiting time for data preparation while optimizing execution and memory use.       

When data is loaded, it is frequently saved in memory before being transferred to the GPU for training via the NCCL interface. Pinged memory, a unique kind of memory allocation that permits DMA transfers between the CPU and GPU, is used by zero-copy data loaders to increase efficiency. By doing this, the overhead of transporting data is decreased, guaranteeing that the GPU gets it fast. Additionally, a queue of batches is kept ready before the model requires them by using data prefetching. Training can continue uninterrupted and resource usage is maximized by overlapping data loading and calculation. Each worker in the pipeline processes a fraction of the data while working under the radical-pilot process management. Each worker operates as a distinct process when multiprocessing is enabled, therefore memory sharing is not the default setting. 
%However, shared memory techniques can be implemented to avoid unnecessary duplication of data across processes, improving memory efficiency. In some cases, threading is used instead of multiprocessing, but due to Python’s Global Interpreter Lock (GIL), threading is generally less efficient for CPU-intensive tasks like data loading.

Effective memory management is a key feature of Deep RC when working with big datasets. Frameworks such as PyTorch's \texttt{DistributedSampler} make sure that each GPU receives a distinct slice of the dataset when many GPUs are being used, avoiding redundancy. Batch sizes also need to be carefully considered; although lower batch sizes may use less memory but result in more updates per epoch, bigger batch sizes speed up training but use more memory. Training performance can be greatly impacted by properly adjusting the batch size, memory settings, and number of workers.

\begin{figure}[htpb]
    \begin{center}
        \includesvg[width=0.95\linewidth]{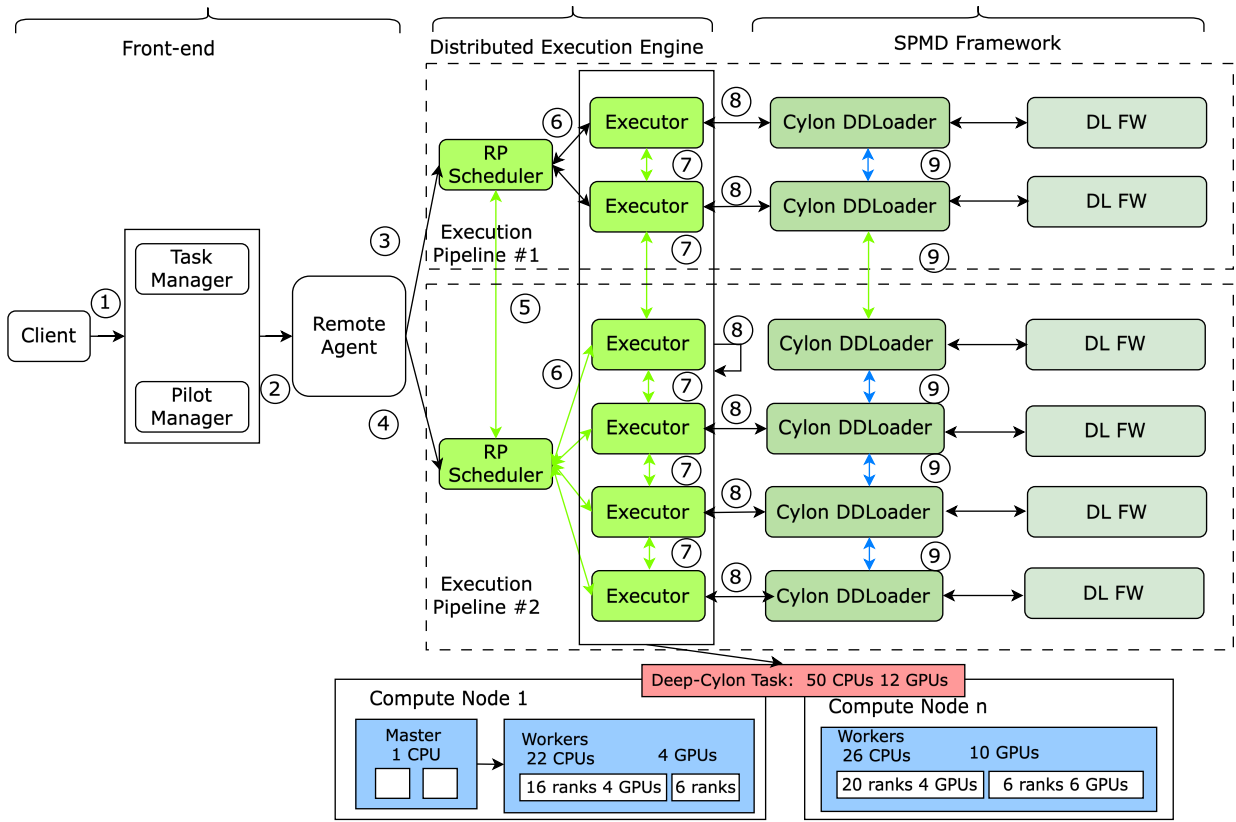}
        \end{center}
        \caption{Deep RC Workflow Architecture. From the bottom-up view, the compute node is a Hardware layer that is compatible with vendor-based CPUs and GPUs. Deep Radical-Cylon(RC) tasks consist of data engineering and deep learning jobs executed on multiple execution pipelines.}
    \label{fig:deepcylon-arch}
\end{figure}

In Fig.-~\ref{fig:deepcylon-arch}, Cylon is connected as a top-level client program to submit different Deep RC tasks (services, functions, or executables) to RADICAL-Pilot for multi-node (CPUs and GPUs) execution. It follows master-workers architectures, where a number of worker nodes are managed by one or more masters, each of whom may have one or more central processing units. 

%In the initial step (Step 1), when a user intends to
Let's now execute a program that calls the RP-Client(1) and does many computations using the Deep RC system. The Pilot Manager then assigns virtual devices for calculations that have not yet been performed and registers these calculations with the Resource Manager (2). After that, the RP-Client gives the background server instructions to execute the pilot manager's commands, accounting for various calculations, including device network connections and data routing (3,4). If a program's virtual device remains constant, the generated representation can be reused fast. However, if the Resource Manager changes the virtual device of a program, recompilation is necessary. Together, these three stages comprise Deep RC's front end. Remote Agent creates several execution pipelines with two persistent daemons, an executor and a scheduler, that can communicate in order to enable distributed coordination, which is the control plane communication (5, 6, 7). The executor invokes the Deep RC bridge to perform local sorting, joining, or deep learning inferencing operations and to construct a distributed data loader(DDLoader) (8). In this scenario, most data plane connections consist of cluster communications involving shuffle, gather, or gradient-sharing activities (9) within the deep learning framework(DL FW). It is noteworthy that the data plane uses the same communication structure; a green arrow indicates lower bandwidth and a blue arrow indicates higher bandwidth.

The main channel of communication between the two systems is through their native APIs because they both provide simple, flexible Python-based interfaces~\cite{de_dynamio_2007}. Both systems are rapidly changing in terms of flexibility, which could result in new, substantial changes to the design or implementation, including the addition or removal of new system components. It is possible to isolate and contain any malfunctioning component, preventing the rest of the system from receiving or carrying out duties.

%The main communication point between both systems is their native APIs, as both systems offer flexible and simple Python-based interfaces ~\cite{de_dynamio_2007}. Flexibility-wise, both systems are developing rapidly and might introduce new fundamental changes in the design or implementation, such as adding or removing new system components. 
%Further, any changes in both systems do not necessarily require changes to the other system's components and would not affect the existing integration as there are no direct dependencies between the integrated systems. 
%From a fault tolerance perspective, the integration approach of Cylon and RADICAL-Pilot is more resilient, as failures in one system or component do not affect the entire system. 
%Failure of any component can be isolated and contained, allowing the rest of the system to continue receiving and executing tasks.

\section{Experiments}

\subsection{Experimental Setup}\label{subsec:exp-setup}
We set up pipeline and scalability studies using UVA Rivanna HPC~\cite{rivanna_2023}. We utilize the \texttt{parallel} queue on Rivanna, which has a maximum of 16 nodes and 40 cores per node.  While performing model training and prediction operations with many pipeline executions, we assess Deep RC's efficiency and contrast it with Bare Metal deep learning. Since Deep RC smoothly supports both PyTorch and Tensorflow, we test the Tensorflow pipeline using the Hydrology Model \cite{he2024science} and the PyTorch pipeline using 11 models from Neuralforecast \cite{olivares2022library_neuralforecast}.  In addition to Total Execution Time and Deep RC overheads from the systems, we measure a number of other parameters. The Total Execution Time shows how long Deep RC took to complete the training or prediction tasks using the N-rank computational resources. Time spent deserializing the task object, building the \texttt{NCCL/GLOO-Communicator} with N ranks, and delivering it to the tasks is represented by the overheads for Deep RC (mostly RP). 
Together, the experiments enable us to compare Deep RC's scalability performance to that of Bare Metal(BM) deep learning on Rivanna using a variety of configurations.

\begin{figure}[htpb]
    \begin{center}
    \includesvg[width=0.7\linewidth]{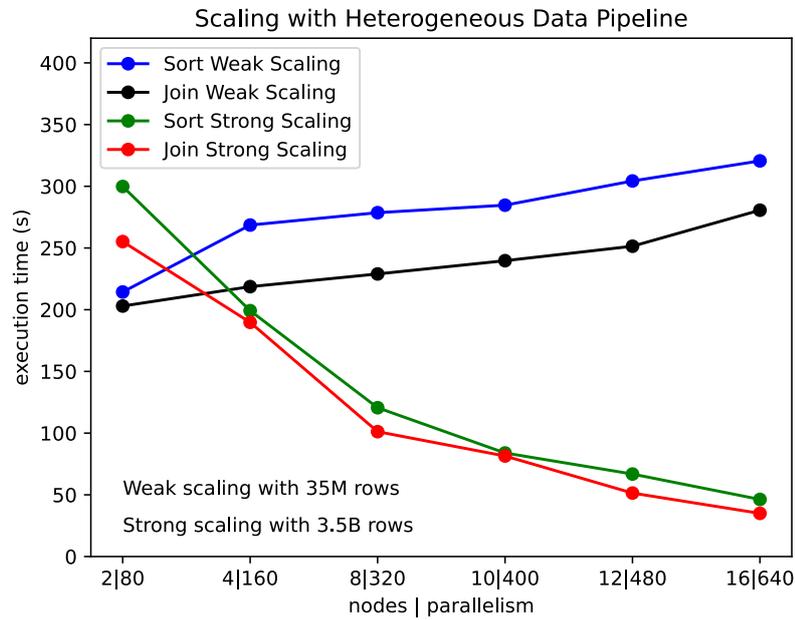}
    \end{center}
    \caption{Heterogeneous Executions with sort and join strong and weak scaling(4) operations on \textbf{\textit{rivanna}}. Strong scaling with 640 parallelism takes a bit more time due to the lack of rows available for each worker and some workers go idle.}
    \label{fig:heterogeneous-scaling}
\end{figure}

%For each task, we ensured the allocation of optimal resources per node. %In the parallel partition of Rivanna, we harnessed up to 37 cores per %node and utilized a maximum of 35 million rows per worker for the %sorting and joining weak scaling.

\subsection{Multiple Data Pipeline scaling operations}\label{subsec:mul_data}

We carried out four distinct scaling procedures as part of the data pretreatment in Fig-\ref{fig:heterogeneous-scaling} in order to demonstrate a heterogeneous data pipeline. We have shown that both strong and weak scaling operations exhibit the predicted tendencies. When number of worker increase, total execution time should decrease despite additional communication overheads across multiple workers in strong scaling. For weak scaling, the same amount of data is allocated to all workers, who are expected to finish the job with additional communication overheads. It is clearly evident that the system can perform all scaling operations on multiple pipeline with expected behavior. The output of all operations will create a Cylon Global Table(GT) and Dataframe which will be used as input for deep learning model training and inferencing in the downstream task. 

\subsection{Pipeline Testing with TensorFlow based Hydrology Model}\label{subsec:exp-hydrology}

%In an ideal scenario, the execution time for each job would remain constant, forming a flat line. However, due to system overhead, total execution time based on 10 iterations is considered for each task. 

\begin{figure}[htbp]
    \includegraphics[width=0.95\linewidth]{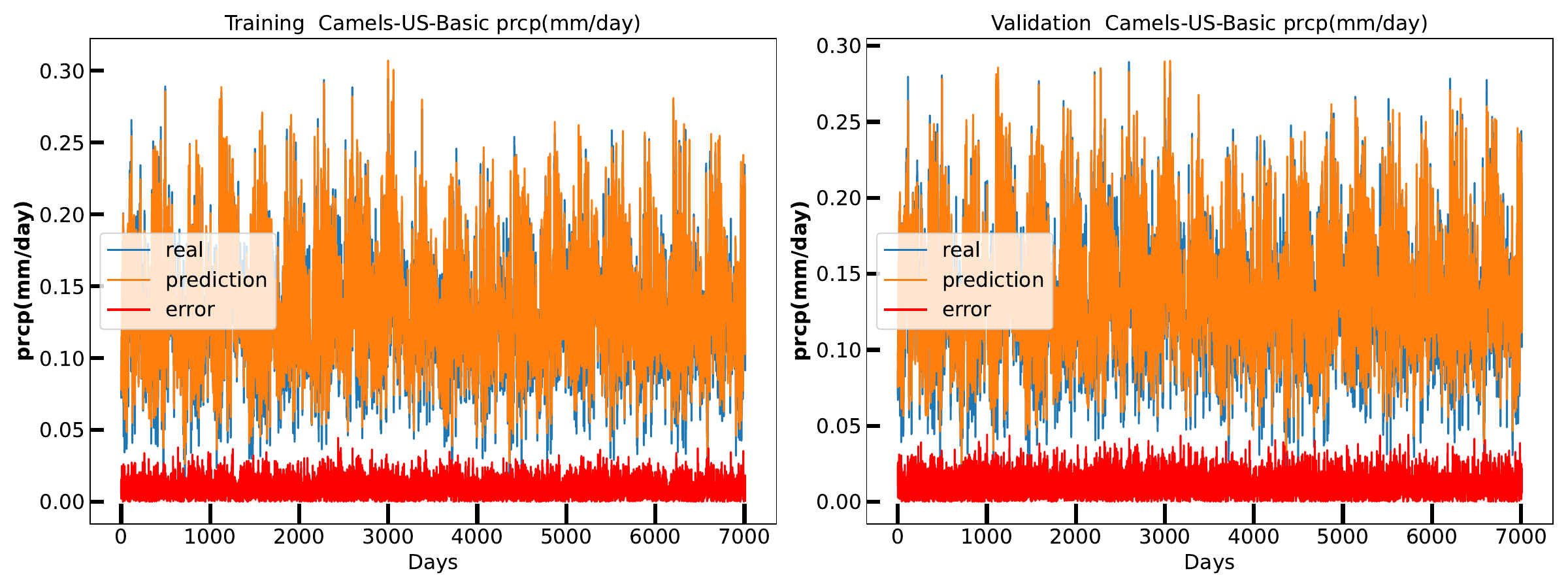}
    \caption{Training(left) and Prediction(right) accuracy of Precipitation with Camels-US datasets \cite{addor2017camels} in LSTM Hydrology model.}
    \label{fig:rivanna-precp-srad1}
\end{figure}

Applying a heterogeneous pipeline to inferencing and model training, however, adds complexity to data loading and gradient gathering over several nodes. We have got our input data from the section \ref{subsec:mul_data}. Due to ongoing RP overheads, the overall execution time for both bare-metal and Deep RC executions on Rivanna varies between 4 and 6 hours. All tests are run four times with different parallelisms (a single rank is used for each parallel execution).  The experimental findings demonstrate how well the LSTM-based hydrology model captures and accurately predicts important factors. The potential of using deep learning pipelines to advance hydrological research and management is demonstrated by its performance.

\begin{figure}[htpb]
    %\begin{center}
    \includegraphics[width=0.95\linewidth]{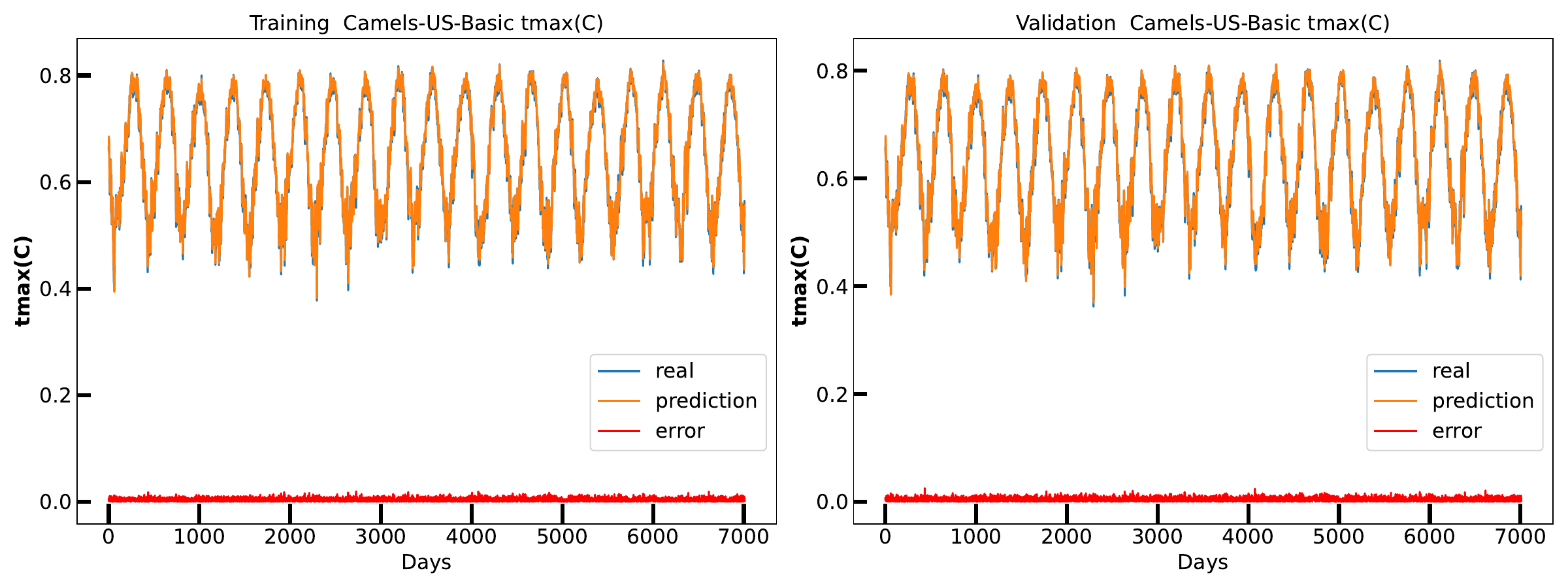}
    \caption{Training(left) and Prediction(right) accuracy of Maximum Temperature with Camels-US datasets in LSTM Hydrology model.}
    %\end{center}
    \label{fig:rivanna-tmp}
\end{figure}

\begin{table}[htpb]
	\centering
	\caption{Performance summary with CAMELS US data. Summed time series MSE: CAMELS-US training MSE = 0.004111, validation MSE = 0.004203%\mtnote{We have to add the data of Summit, otherwise the reviewers will complain that the presentation is partial. As we are very tight with space, you could span the table on two columns (use the star) and then put weak and strong as columns instead of as rows. After that, add a new table for Summit as the parallelism is different.}
 }
	\label{tab:a-b-exp_table1}
	\begin{tabular}{llcr @{\hspace{1\tabcolsep}} rr @{\hspace{1\tabcolsep}} c}
		\toprule
            \multicolumn{1}{l}{}         &
            		                              &
            \multicolumn{1}{c}{}         &
		\multicolumn{2}{c}{}  \\
		  Properties             &
		Execution                      &
		MSE                     &
		\multicolumn{2}{c}{NNSE}                    &
	    \multicolumn{2}{c}{Execution Time(s)}                    \\
		\midrule
		\multirow{2}{*}{Precipitation} &
		\multirow{1}{*}{Train} &
		0.003508    & &
		0.820     & & 
		          \\
		\cmidrule{2-6}
		                    & 
            \multirow{1}{*}{Val} &
		0.003585   & &
		0.819     & & 
		    \\
		\cmidrule{1-6}
		%
		% \multirow{3}{*}{2}      &
		\multirow{1}{*}{Mean} &
		\multirow{1}{*}{Train} &
		0.000276                  & &
		0.961         & &
		6482.24(Train)  \\
		\cmidrule{2-6}
		\multirow{1}{*}{Temperature} &
		\multirow{1}{*}{Val} &
		0.000283                     & &
		0.960     & &
		1927.70(Val)  \\

		\cmidrule{1-6}
		%
		% \multirow{3}{*}{2}      &
		\multirow{2}{*}{Streamflow} &
		\multirow{1}{*}{Train} &
		0.000287            & &
		0.806     & &
		  \\
		\cmidrule{2-6}
		  &
		\multirow{1}{*}{Val} &
		0.000296                & &
		0.812 & &
		 \\

		\bottomrule
	\end{tabular}
\end{table}

\begin{table}
	\centering
	\caption{Deep RC Execution Time and RP Overheads of different Neural Forecast and Hydrology models with Training Task on Rivanna. %\mtnote{We have to add the data of Summit, otherwise the reviewers will complain that the presentation is partial. As we are very tight with space, you could span the table on two columns (use the star) and then put weak and strong as columns instead of as rows. After that, add a new table for Summit as the parallelism is different.}
 }
	\label{tab:a-b-exp_table2}
	\begin{tabular}{llcr @{\hspace{1\tabcolsep}} lr @{\hspace{1\tabcolsep}} l}
		\toprule
            \multicolumn{1}{l}{Forecasting}         &
            		                              &
            \multicolumn{1}{c}{GPUs}         &
		\multicolumn{2}{c}{Execution Time}         &
	    \multicolumn{2}{c}{Overheads}              \\
		  Domain             &
		Model               &
		a100 80GB                     &
		\multicolumn{2}{c}{time (seconds)}                    &
	    \multicolumn{2}{c}{(tasks/second)}                    \\
		\midrule
		\multirow{2}{*}{Hydrology} &
		\multirow{2}{*}{LSTM} &
		1                       &
		14456.64 & $\pm4.97$     &
		4.13 & $\pm1.1$          \\
		                    & &
		2                     &
		10216.52 & $\pm4.26$      &
		4.13 & $\pm1.6$        \\
            		        & &
		4                     &
		6482.24 & $\pm5.13$      &
		5.01 & $\pm1.82$        \\
		\midrule
		%
		% \multirow{3}{*}{2}      &
		\multirow{4}{*}{Neural} &
		\multirow{1}{*}{Autoformer} &
		2                         &
		189.15 & $\pm3.23$         &
		3.56 & $\pm0.8$              \\
		\multirow{4}{*}{Forecast} &
		\multirow{1}{*}{AutoNHITS} &
		2                      &
		500.8 & $\pm3.13$        &
		3.51 & $\pm0.41$              \\
		&
		\multirow{1}{*}{TFT} &
		2                      &
		298.13 & $\pm3.23$        &
		3.12 & $\pm0.9$            \\

	         &
		\multirow{1}{*}{TimesNet} &
		2                      &
		806.09 & $\pm4.84$         &
		4.69 & $\pm0.21$          \\

		&
            \multirow{1}{*}{DeepAR} &
		2                     &
		72.32 & $\pm3.35$        &
		3.45 & $\pm0.51$          \\

		&
            \multirow{1}{*}{VanillaTransformer} &
		2                    &
		269.18 & $\pm4.67$          &
		4.56 & $\pm0.33$        \\
		\bottomrule
	\end{tabular}
\end{table}

\begin{figure}[htpb]
    \begin{center}
    \includesvg[width=0.7\linewidth]{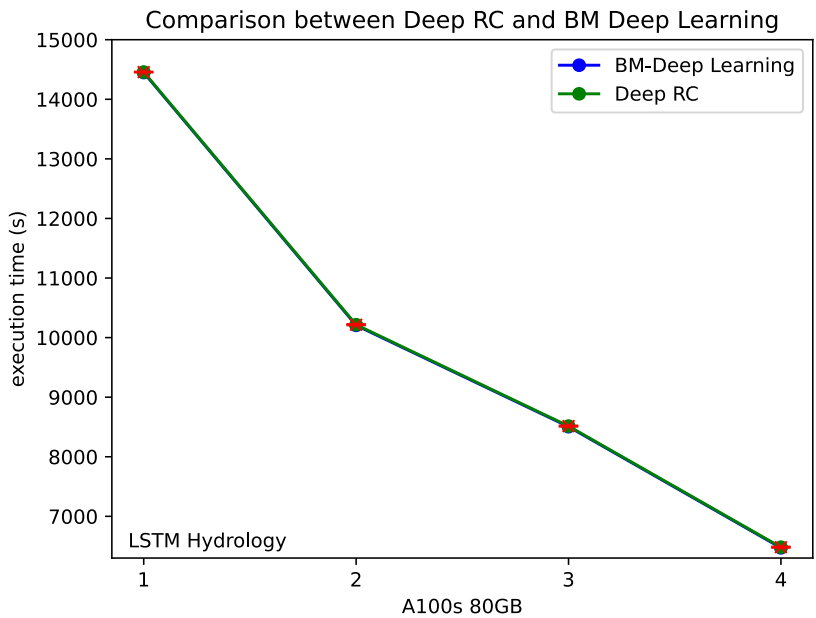}
    \caption{Execution time comparison. Radical-pilot overhead is so negligible that BM Deep Learning and Deep RC graphs overlap. We used 4 different execution configurations to train the LSTM Hydrology model. BM Deep Learning execution are 14448.81, 10205.37, 8504.53, 6471.71 where Deep RC execution time is shown in Table-~\ref{tab:a-b-exp_table1}}
    \label{fig:rivanna-cmp-res}
    \end{center}
\end{figure}

\begin{figure}[htbp]
    \includegraphics[width=0.9\linewidth]{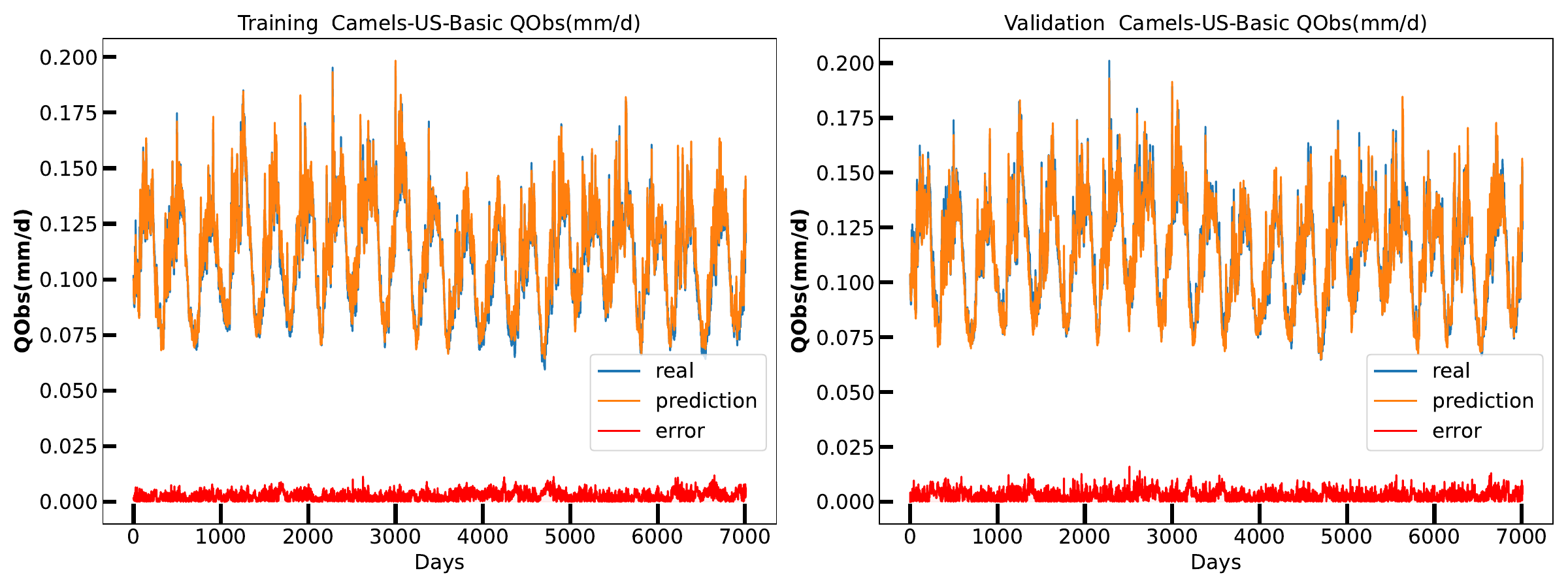}
    \caption{Training(left) and Prediction(right) accuracy of Observed Streamflow(QObs) with Camels-US datasets in LSTM Hydrology model.}
    \label{fig:rivanna-vp-qobs2}
\end{figure}

Key hydrological parameters were successfully predicted by the LSTM-based hydrology model, which achieved a training MSE range of 0.000276 - 0.003508 and a prediction error range of 0.000283 - 0.003585 in Mean Temperature, Streamflow and Precipitation shown in Table- \ref{tab:a-b-exp_table1}. The model's capacity to efficiently learn and generalize temporal patterns in the dataset is demonstrated by these low error values. The model's exceptional accuracy in capturing the relationships between temperature (Fig.- \ref{fig:rivanna-tmp}), precipitation(Fig.-\ref{fig:rivanna-precp-srad1}), and streamflow(Fig.-\ref{fig:rivanna-vp-qobs2}) during training suggests that the LSTM architecture is a good fit for hydrological forecasting.

The prediction performance was particularly outstanding, with errors remaining continuously low over the test dataset. Temperature forecasts showed minimal variance from observed values, showing the model's effectiveness in capturing reasonably smooth temporal trends with minimal execution cost. In Table-\ref{tab:a-b-exp_table2}, we have shown model training time with multiple configurations to prove the performance of the system. We see constant overheads between 6 to 8 seconds compared to a total time of 6482.24 to 14456.64 seconds which is very negligible. We performed 11 concurrent validations and shown them in Table-\ref{tab:a-b-exp_table1}, column Execution Time(s), and got 1927.70 seconds which is very reasonable, and got similar results(Fig.-\ref{fig:rivanna-cmp-res}) compared to the Hydrology baseline model \cite{he2024science}.

\begin{figure}[htbp]
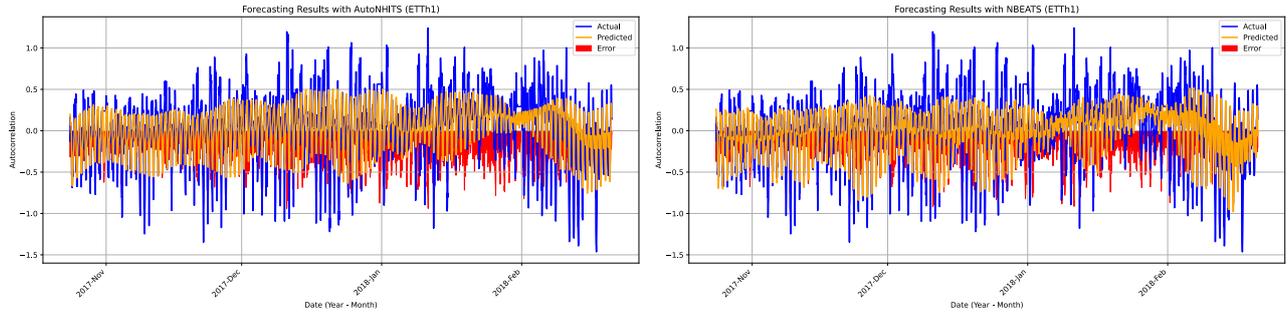

    \centering
    % First Figure (4 model pairs)
    \includesvg[width=0.5\linewidth]{Figure/AutoNHITS_ETTh1_forecast}\hfill
    \includesvg[width=0.5\linewidth]{Figure/NBEATS_ETTh1_forecast}\hfill
    \caption{Training and Prediction accuracy of electrical transformers' oil temperature for AutoNHITS and NBEATS model. The dataset used is the Electricity Transformer Dataset (ETDataset) \cite{haoyietal-informer-2021}}
    \label{fig:vp-qobs-1}
\end{figure}

\subsection{Pipeline Testing with PyTorch based Multiple forecasting Model}\label{subsec:exp-sort}
Deep RC has the capability to perform deep learning jobs with PyTorch-based models. We choose 11 Neuralforecast models for training and prediction jobs. After training 11 models, the prediction job is run 10 times for each model(total of 110 tasks) as a parallel function. Each parallel execution on Rivanna uses a single rank, and the same scaling configurations are used for the training and inference processes of neural forecast models. We got the expected precision in all models and generated the results with 3 metrics(MAPE, MAE, MSE) shown in Table-\ref{tab:non_raptor_vs_raptor} which are similar to the baseline execution shown in the Neuralforecast experiment\cite{olivares2022library_neuralforecast}. We have shown training and prediction graphs(Fig.-\ref{fig:vp-qobs-1}) for 2 models(AutoNHITS and NBEATS) due to limited space. All results are archived to the github repository.   

%Higher rank numbers are expected to cause a latency increase since they affect the data shuffle and merging steps of the distributed process as well as model training, adding extra overhead. A key factor affecting execution time is the efficient use of resources for data partitioning and communication. An overlapping error bar that shows comparable performance with both measures is what we are obtaining after several revisions. Execution time can be decreased by dividing a large dataset over many nodes. 

Although the total execution times of the Deep RC and BM Deep Learning approaches differ by 1 to 5 seconds, training time is less and does not impact on prediction process. Aside from the comparable performance, we observe a constant overhead when using Deep RC in strong scaling operations despite increasing parallelism. Distributed execution presents a set of challenges that include managing data distribution, navigating communication overhead between nodes, and mitigating potential node failures, which are magnified with an increased number of nodes.

\begin{table}[htbp]
\centering
\caption{Single pipeline testing by measuring Mean Absolute Error(MAE), Mean Squared Error(MSE), and Mean Absolute Percentage Error(MAPE) to ensure model training works as expected. Comparison of training time with Deep RC and Bare Metal Deep Learning(BM DL) with 400 epochs. We observe constant overheads(approximately 4.15 seconds on average) in a single pipeline.}
\label{tab:non_raptor_vs_raptor}
\begin{tabular}{lc|ccc|c}
\hline
\multicolumn{1}{c}{\textbf{Model}} 
 & \textbf{Train(s)} 
 & \textbf{MAE} 
 & \textbf{MSE} 
 & \textbf{MAPE(\%)} 
 & \textbf{Train(s)} \\
 %& \textbf{MAE} 
 %& \textbf{RMSE} 
 %& \textbf{MAPE (\%)} \\
\multicolumn{1}{c}{} & \multicolumn{1}{c}{\textbf{BM DL}} & \multicolumn{4}{r}{\textbf{Deep RC}} \\
\hline
% ---------------- Figure 1 models ----------------
\textbf{Autoformer}        & 185.51     & 0.51    & 0.57    & 2.65   & 189.15  \\
\textbf{DeepAR}            & 72.14     & 0.50    & 0.59    & 2.48   & 76.32  \\
\textbf{NLinear}         & 20.98     & 0.42    & 0.39    & 2.45   & 24.34  \\
\textbf{GRU}               & 157.15     & 0.52     & 0.59     & 2.18    & 163.2    \\
% ---------------- Figure 2 models ----------------
\textbf{NBEATS}            & 15.21     & 0.39        & 0.36        & 2.01    & 19.03    \\
\textbf{AutoNHITS}         & 495.78    & 0.39        & 0.33        & 2.22    & 500.8   \\
\textbf{PatchTST}          & 47.07     & 0.37    & 0.32    & 2.20   & 51.15    \\
\textbf{TFT}               & 293.79     & 0.42     & 0.47     & 1.28    & 298.13     \\
% ---------------- Figure 3 models ----------------
\textbf{TimesNet}          & 801.13     & 0.39     & 0.36     & 2.51    & 806.09    \\
\textbf{VanillaTransformer}& 264.57     & 0.46    & 0.51    & 1.43   & 269.18    \\
% -------------- Extra model (not in figures) ------
\textbf{TiDE}      & 19.39     & 0.36     & 0.34     & 1.84    & 23.61   \\
\hline
\end{tabular}
\end{table}

\begin{table}[htbp]
\centering
\caption{Performance Comparison of multiple pipelines with Deep RC and Bare Metal Deep Learning(BM DL) with PyTorch and TensorFlow based models, runs on 2 a100s 80GB GPUs. For cylon job, it uses 8 nodes with 40 cores/node}
\label{tab:pytorch_tensorflow}
\begin{tabular}{lccc|c|c}
\hline
\multicolumn{1}{c}{\textbf{Pipeline}} 
 & \textbf{Number of} 
 & \textbf{Cylon} 
 & \textbf{DL} 
 & \textbf{BM-DL} 
 & \textbf{Deep RC} \\
 %& \textbf{MAE} 
 %& \textbf{RMSE} 
 %& \textbf{MAPE (\%)} \\
 \multicolumn{1}{c}{\textbf{Type}} 
 & \textbf{pipelines} 
 & \textbf{Task} 
 & \textbf{Task} 
 & \textbf{Execution(s)} 
 & \textbf{Execution(s)} \\
 
%\multicolumn{1}{c}{} & \multicolumn{1}{c}{\textbf{Non-Raptor}} & \multicolumn{4}{r}{\textbf{Raptor}} \\
\hline
% ---------------- Figure 1 models ----------------
\textbf{Hydrology}        & 11     & join    & inferencing    & 21381.73135   & 21305.83772  \\
\textbf{NeuralForecast}   & 11     & join    & inferencing   &  167.8454124  & 164.5677196  \\
\hline
\end{tabular}
\end{table}

\subsection{Discussions}
\label{sec:discussions}
The results show that Deep RC maintains comparable speed in multi-task execution while achieving effective scaling with low overhead. Furthermore, during heterogeneous execution, it outperforms the BM-Deep Learning model's batch processing capabilities. We have developed 11 pipelines with one Cylon join and 11 deep learning inferencing jobs using an LSTM-based Hydrology and NeuralForecast model in Table- \ref{tab:pytorch_tensorflow}. It significantly decreased 75.9  and 3.28 seconds in both experiments, which is important for inferencing tasks. Because any commercial cloud platform that manages thousands of requests would be significantly impacted. In comparison to the overall execution time and the scale of trials, Radical-Cylon consistently generates a ~\texttt{MPI/NCCL/GLOO-Communicator} with numerous ranks in constant time, according to the overheads associated with Deep RC. However, when we attempted to infer LLMs, the RP scheduler and resource allocation module—had trouble assigning resources for Deep RC. To manage such jobs, a design modification incorporating multi-level parallelism is necessary. This significant design modification will be presented separately because we view it as a future work.  

%The performance of Cylon is evaluated through data frame execution runtime, where data frame operators are organized into a directed acyclic graph (DAG). Execution efficiency can be enhanced by identifying independent branches within the DAG and running those tasks in parallel. Each task follows a Bulk Synchronous Parallel (BSP) approach, and Deep Radical-Cylon enables precise control over BSP task parallelism. Future improvements to Cylon include integrating an optimizer for data frame DAGs, focusing on traditional query optimization akin to SQL and addressing scheduling overheads within the broader scheduling ecosystem. These enhancements aim to optimize data processing for machine and deep learning tasks.

Deep RC was created to accommodate a variety of data pipelines and enable unified execution on both CPUs and GPUs. Despite the computational complexity that comes with combining CPUs and GPUs into a single job, heterogeneous execution is still possible by using different rank groups with memory dedicated to either CPUs or GPUs. The initial focus of Deep RC, has been on setting up several data pipelines and using functions to carry out distributed operations; Cylon and deep learning frameworks are essential to making distributed execution possible. As multi-tenancy scenarios get more complicated, RADICAL-Pilot will need to handle a variety of resource types, including CPUs/GPUs allocation and host or device memory. Large-scale resource allocation and reliable monitoring are features of the master-worker paradigm. We plan to support all multi-tenancy requirements, e.g. prioritization, performance isolation, and resource tracking in the future. Compared to earlier efforts in Radical-Cylon, Deep RC seeks to produce scalable and effective deep learning pipelines on much bigger resource pools in a shorter amount of time.

\section{Related Works} 
\label{sec:relatedWorks}

In order to increase the scalability, effectiveness, and performance of large-scale AI systems, recent developments in distributed computing and deep learning research have placed an increasing emphasis on novel frameworks and architectures. The Ray framework, for instance, has recently been used by researchers to streamline workflows for distributed reinforcement learning \cite{moritz2018ray}. Distributed policy assessment in high-dimensional contexts was implemented using Ray. Compared to conventional MPI-based configurations, the framework's capacity to handle diverse workloads across clusters allowed for a reduction in training time. Another example is the use of Google Research's Pathways architecture into state-of-the-art multimodal AI models. Complex models that needed to analyze text, pictures, and temporal input simultaneously were trained using Pathways. By combining data from many SPMD units, researchers were able to reduce hardware usage and accomplish smooth task coordination by utilizing Pathways' MPMD design\cite{barham2022pathways}. 

Similarly, current AI research has relied heavily on Apache Spark\cite{zaharia2016apache} and Apache Flink\cite{carbone2015apache} to process large datasets. Terabytes of genomic data are preprocessed using Spark's in-memory processing capabilities to enable deep learning-based analysis later on. Flink's ability to handle continuous data streams was demonstrated when its stream-processing capabilities were used in real-time anomaly detection systems for industrial Internet of Things applications. Frameworks like Dask \cite{rocklin2015dask} have been widely used in the data analytics field. A gradient-boosted decision tree model's hyperparameter tweaking computational load is divided among GPU clusters using Dask.

CuDF has emerged as a key technique for GPU-accelerated ETL pipelines. CuDF was utilized by researchers to integrate high-resolution video feeds with preprocessed 2D/3D data. Because of this configuration, preprocessing latency was decreased, allowing multi-modal and other decision-making models to infer in real time \cite{sarker2022incremental, sakib2024electronic}. 
%CuDF integrates seamlessly with distributed computing frameworks like Dask and RAPIDS, enabling the creation of distributed data loaders that scale across multiple nodes in a cluster. 
%The system as a whole may preprocess data in parallel since each node handles a portion of the dataset using its local GPU. This increases system throughput overall and reduces communication overhead. Data transfers between the CPU and GPU are reduced when CuDF is used to process data directly on the GPU. In distributed systems, where frequent data transfers can result in significant latency, this has a particularly significant impact. 
%CuDF lessens these delays and creates a more effective data pipeline by maintaining data on the GPU during the preprocessing and loading phase. 
CuDF facilitates effective handoff between data loaders and model training pipelines through its interface with frameworks such as TensorFlow, PyTorch, and RAPIDS.ai. There is no need for extra transformations or data copying processes when converting CuDF-prepared data straight into GPU tensors. Even in high-throughput situations, this close integration makes sure that the data flow doesn't become a bottleneck. Researchers used CuDF to preprocess textual input in a recent work on training big language models, which included batching, tokenization, and normalization. The preprocessing speed rose by up to 15x when these jobs were distributed across a GPU cluster using Dask-CuDF, and the training pipeline maintained near-linear scalability as the dataset size increased \cite{daskcudf:online}. Even though CuDF has many benefits, there are still issues in integrating it smoothly in environments that use both CPU and GPU resources. CuDF's compatibility with these hybrid systems is being improved through continuing research, opening the door to even more reliable distributed data loading pipelines.

Frameworks designed specifically for deep learning, such as OneFlow, have also drawn interest. The potential for OneFlow to take the place of conventional workflows based on TensorFlow in large-scale transformer model training \cite{yuan2021oneflow}. In order to achieve greater scalability, the researchers emphasized OneFlow's simultaneous processing and efficient memory management. In studies on hybrid distributed systems, ZeroMQ has been investigated at the communication level. ZeroMQ was utilized in a 2024 study to lower communication overhead and latency among participating nodes by enabling asynchronous communication in federated learning configurations \cite{zmq1:online}. Similar to this, a recent effort that trained generative models on decentralized datasets made use of technologies like Parsl, which enabled better resource allocation and dynamic task scheduling \cite{babuji2019parsl}.

These papers highlight how important distributed frameworks and architectures are to the advancement of AI and deep learning research. Researchers are pushing the limits of what is feasible in scaled AI systems by combining technologies such as Ray, Pathways, Spark, Flink, and OneFlow. Using Deep RC, we shade a portion of the deep learning execution.

\section{Conclusions}
\label{sec:conclusion}

In conclusion, Deep RC successfully attains performance parity with state-of-the-art multi-execution designs, confirming its efficacy in the rapidly changing fields of deep learning execution pipelines and data engineering. It offers a unified framework that simplifies model training, prediction, and data processing under a centralized execution paradigm by tackling the challenges of resource management and varied pipeline execution. Through this change, task management becomes more organized and scalable while also improving computing efficiency. Our analysis demonstrates Radical Cylon's versatility across a range of distributed tasks, including the capacity to smoothly interleave client workloads, optimize deep learning execution pipelines, and reduce computational overhead. Several advanced cluster management rules, such as virtualization and multi-execution sharing, can be reimplemented to meet the particular requirements of machine learning and big data workloads in the future. These developments establish Deep RC as a strong and adaptable solution that can spur innovation in deep learning and large-scale data processing ecosystems.

\section*{Acknowledgments}
We gratefully acknowledge the support from the Department of Energy and National Science Foundation through DE-SC0023452, NSF 1931512,NSF 2103986, and OAC-2411009 grants. The RADICAL team thanks Andre Merzky and Matteo Turilli for their support and development of RADICAL-Cybertools.

\begin{comment}
\section{Double blind review}
\newcommand{\Cylon}{\censor{Cylon}}
\newcommand{\Radical}{\censor{Radical Pilot}}

Gregor: They changed to double-blind review. 
I put paper anonymous-acm.pdf. 

This is \Cylon.
This is \Radical.

This is text blocked out \censor{secret}

Use bibtex for references 
\end{comment}

\bibliographystyle{splncs04}
\bibliography{conference_101719}

\end{document}